\documentclass[%
 reprint,
 superscriptaddress,
 amsmath,amssymb,
 aps,
 pre,
]{revtex4-1}

\usepackage{graphicx}% Include figure files
\usepackage{dcolumn}% Align table columns on decimal point
\usepackage{bm}% bold math
\usepackage[normalem]{ulem}
\begin{document}

\pagestyle{empty}

\title{Cooperativity transitions driven by higher-order oligomer formations in \\
ligand-induced receptor dimerization}

\author{Masaki Watabe}
\thanks{Corresponding author}
\email{masaki@riken.jp}
\affiliation{Laboratory for Biologically Inspired Computing, RIKEN Center for Biosystems Dynamics Research, Suita, Osaka 565-0874, Japan}
\author{Satya N. V. Arjunan}
\affiliation{Laboratory for Biologically Inspired Computing, RIKEN Center for Biosystems Dynamics Research, Suita, Osaka 565-0874, Japan}
\affiliation{Lowy Cancer Research Centre, The University of New South Wales, Sydney, Australia}
\author{Wei Xiang Chew}
\affiliation{Laboratory for Biologically Inspired Computing, RIKEN Center for Biosystems Dynamics Research, Suita, Osaka 565-0874, Japan}
\affiliation{Physics Department, Faculty of Science, University of Malaya, Kuala Lumpur 50603, Malaysia}
\author{Kazunari Kaizu}
\affiliation{Laboratory for Biologically Inspired Computing, RIKEN Center for Biosystems Dynamics Research, Suita, Osaka 565-0874, Japan}
\author{Koichi Takahashi}
\thanks{Corresponding author}
\email{ktakahashi@riken.jp}
\affiliation{Laboratory for Biologically Inspired Computing, RIKEN Center for Biosystems Dynamics Research, Suita, Osaka 565-0874, Japan}
\affiliation{Institute for Advanced Biosciences, Keio University, Fujisawa, Kanagawa 252-8520, Japan}
\affiliation{Department of Biosciences and Informatics, Keio University, Yokohama, Kanagawa 223-8522, Japan}

%             %  but any date may be explicitly specified
%\date{November 27, 2019}% It is always \today, today,

\begin{abstract}
While cooperativity in ligand-induced receptor dimerization has been linked with receptor-receptor couplings via minimal representations of physical observables, effects arising from higher-order oligomer (e.g., trimer and tetramer) formations of unobserved receptors have received less attention. Here, we propose a dimerization model of ligand-induced receptors in multivalent form representing physical observables under basis vectors of various aggregated receptor-states. Our simulations of multivalent models not only reject Wofsy-Goldstein parameter conditions for cooperativity, but show higher-order oligomer formations can shift cooperativity from positive to negative.
%Cooperativity in ligand-induced receptor dimerization (e.g., EGF receptors) can be linked with receptor-receptor couplings via the state-vector representations of physical observables. The cooperative effects that arise from higher-order oligomer formations of unobserved receptors have, however, received less attention. In this letter, we propose the dimerization model of ligand-induced receptors in the multivalent form that represents physical observables under basis vectors of various aggregated receptor-states. We then evaluate the cooperative responses in the multivalent models, comparing to the cooperativity predicted from the Wofsy-Goldstein formulation. Our results of model simulations not only refute those predictions but also show how higher-order oligomer formations of unobserved receptors in the multivalent models can shift cooperativity from positive to negative.
%A broad range of biological phenomenon is regulated by cooperative binding processes where a large number of biomolecules automatically develop collective behaviors. Understanding what biomolecular network structures can lead to such cooperative responses is of significant interest in various biophysical systems: oxygen transport with hemoglobins, enzyme catalyses and gene regulations.
\begin{description}
\item[DOI] 10.1103/PhysRevE.100.062407
\end{description}
\end{abstract}

%\keywords{cooperative binding, cooperativity, ligand-induced receptor dimerization}%Use showkeys class option if keyword
%                              %display desired
\maketitle

%\tableofcontents

%\section*{Introduction}
\textit{Introduction.} 
Collective behavior is a phenomenon common in human, animal, cellular, and biomolecular systems. Despite varying significantly in terms of the type and composition of biological components, fundamental dynamical properties allow collectives to exhibit rapid or gradual responses in complex environments. For example, group behaviors of wild baboons have been precisely linked to characteristic ``S"-shaped (or sigmoid) response curves via pairwise interactions of subgroups~\cite{strandburg2017, *strandburg2015}. The dynamics of animal groups also exhibit parallels with collective behaviors among biomolecules in living cells, such as the process by which hemoglobin binds to oxygen~\cite{koshland2002, ferrell2009, stefan2013, phillips2013}. Thus, by abstracting structural details of biomolecules and relating biochemical interactions directly to mathematical networks, we can consider key insight regarding collective behavior and likely demonstrate various aspects of biomolecular binding systems.

Collective biomolecular behavior is generally referred to as cooperativity, its main functions being to allow biomolecular binding systems to exhibit either positive or negative sigmoid responses~\cite{koshland2002, ferrell2009, stefan2013, phillips2013}. For example, a conformational change in proximal and distal regions of hemoglobin complex enables efficient transport of oxygen between the lungs and tissue, exhibiting positive cooperativity: steeper sigmoid (or switch-like) responses with a threshold in a concentration range of stimuli~\cite{koshland2002, ferrell2009, stefan2013, phillips2013}. Receptor systems coupling to G-proteins can, however, display more gradual sigmoid curvature, achieving less decisive but also less restricted with respect to a wide concentration range of signaling molecules~\cite{koshland2002, ferrell2009, stefan2013}. Such gradual cooperative responses are known as negative cooperativity. 

%Thus, overall pictures and functionalities of cooperativity in the biomolecular binding systems can vary largely from positive to negative.
%
%its main functions being to allow cellular systems to sense physical and biochemical signal information that acts upon internal or external environments
%Cooperative functions can be characterized into either positive or negative sigmoid responses.
%
%The results of most experimental studies on this topic have shown that the negative cooperative effects in the dimer formations have strongly coupled with direct (or second-order) interactions of two single receptors \cite{wofsy1992a, wofsy1992b, klein2004, mayawala2005a, mayawala2005b, uyemura2005, teramura2006, ozcan2006, macdonald2008, lemmon2008, adak2011, pike2012, hiroshima2012, *hiroshima2013}. In those studies, the simplest dimerization models for equilibrium binding of ligands to receptors were mainly used for fitting to experimental data from living cells. 

In standard systems biology approaches, the fundamental rules governing cooperativity in living cells can be investigated by mapping and analyzing biomolecular networks and their parameter conditions. A key challenge of analyzing network models is finding meaningful and nonintuitive effects. Using data-driven (or inductive) modeling approaches, network models are generally constructed with various biochemical parameters (e.g., equilibrium binding constants) but restricted to observable components (or states) imposed by experimental techniques (e.g., live-cell imaging via biomolecules tagged with fluorescence emitters). The network models can lead to the parameter conditions exhibiting either positive or negative cooperativity. It is often overlooked, however, that these conditions can vary greatly by incorporating realistic but unobserved components (or states) into the network models. For example, the Monod-Wyman-Changeux (MWC) model that describes allosteric regulations of proteins always exhibits positive cooperativity; nevertheless, modifications in the scale of protein's conformational changes proposed by Koshland, Nemethy and Filmer, offer parameter regions that allow negative cooperativity~\cite{monod1965, koshland1966, alan1999, koshland2002}. By introducing hidden components (or states) in the network models, cooperativity can shift from positive to negative (or negative to positive). Such cooperativity transitions thus hinder physical interpretations of the cooperativity extracted from the data-driven modeling approaches.

%
%For example, Monod-Wyman-Changeux (MWC) model can always exhibit positive cooperativity; however, scale modification of a conformational change incorporated into the MWC model (also known as KNL model) offers the possibilities allowing negative cooperativity~\cite{alan1999}.
%
%In the standard approaches of systems biology~\cite{lim2013, chau2012, ma2009}, the origin of such cooperative characteristics in biomolecular binding systems can be generally investigated by mapping and analyzing biomolecular networks and parameter conditions. Extracting the underlying rules that can achieve cooperativity is, however, limited by observable components (or states) imposed on network models. The network models composed of various biochemical parameters such as equilibrium binding constants, can lead to the parameter conditions exhibiting either positive or negative cooperativity. It is often overlooked that these conditions can vary greatly by incorporating ``unobserved" components (or states) into the network models.
%
%The origin of cooperativity in biomolecular binding systems is limited by observable components (or states) imposed on network models. The network models composed of various biochemical parameters (e.g., equilibrium binding constants), lead to the parameter conditions exhibiting either positive or negative cooperativity. It is often overlooked, however, that these conditions can vary greatly by incorporating ``unobserved" components (or states) into the network models.
%

\begin{figure}
\centering
\includegraphics[width=0.98\linewidth]{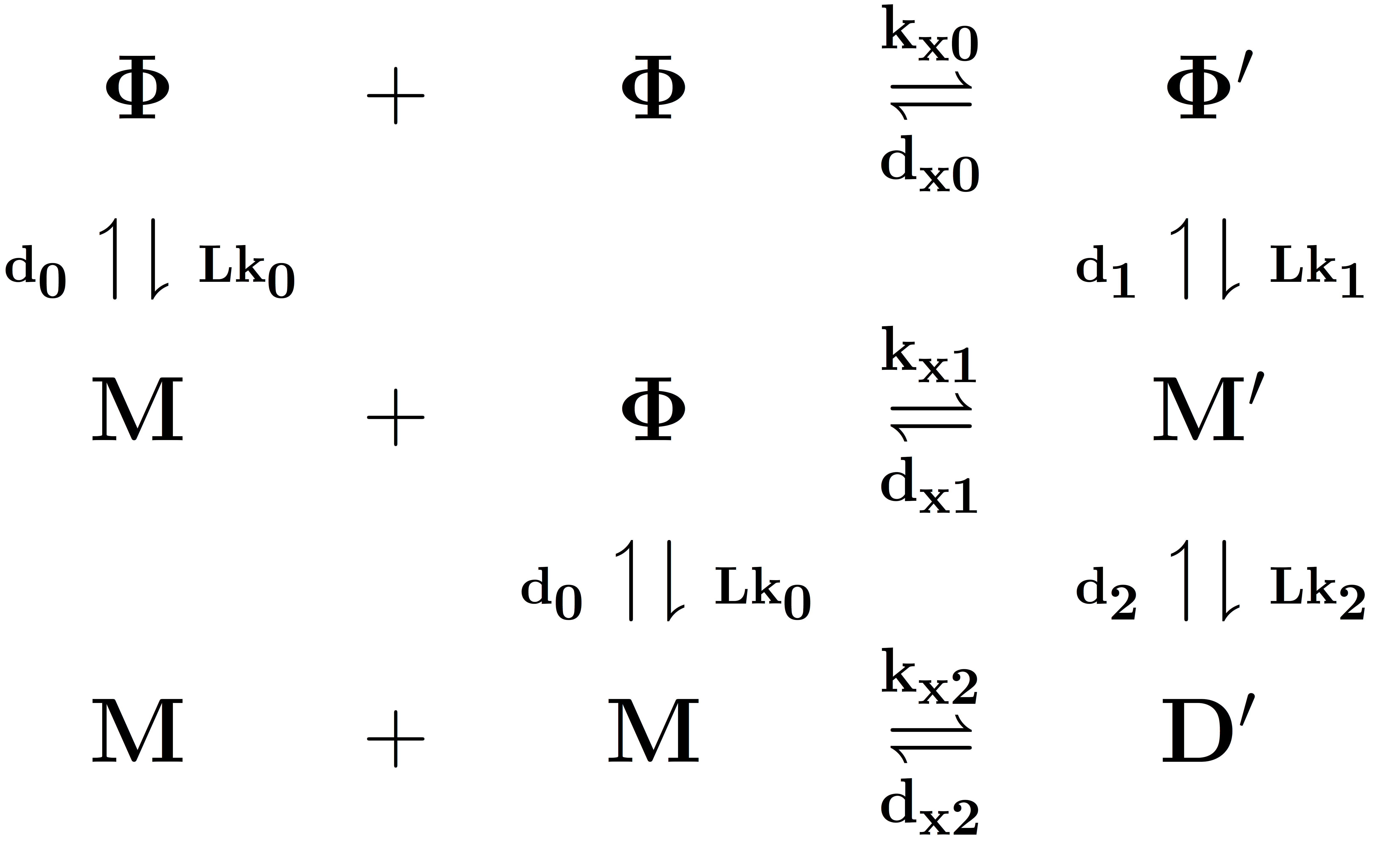}
\caption{{\bf Network model of observed receptor-states in dimer formations}. ${\bf \Phi}$/${\bf \Phi'}$, ${\bf M}$/${\bf M'}$ and ${\bf D'}$ are the observed receptor-state vectors in the formation of nulls, monomers and dimers, respectively. Each observed receptor-states are represented under basis vectors of various aggregated receptor-states. ${\bf k_i}$ and ${\bf d_i}$ are the association and dissociation rates of the $i$-th index, respectively. ${\bf L}$ is the ligand concentration.}
\label{fig01;network}
\end{figure}

Collective behavior of cell-surface receptors is a key function for enabling the efficient transduction of biochemical signals to cellular interiors. Prior experimental studies have explored the origin of negative cooperativity in dimer formations for equilibrium binding of ligands to cell-surface receptors, mainly using the simplest dimerization model formulated by Wofsy and Goldstein~\cite{wofsy1992a, *wofsy1992b, klein2004, mayawala2005a, mayawala2005b, ozcan2006, macdonald2008, lemmon2008, adak2011, pike2012, hiroshima2012, *hiroshima2013}. While this dimerization model predicts the parameter conditions that give rise to negative cooperativity, effects arising from higher-order oligomer (e.g., trimer and tetramer) formations of unobserved receptors have received less attention. In this article, we consider dimerization models in multivalent form that represent physical observables under basis vectors of various aggregated receptor-states (see Figure~\ref{fig01;network}). We then evaluate the cooperative behavior that arises from multivalent models, comparing the cooperativity expected from the Wofsy-Goldstein (WG) formulation. Our results from model simulations imply violation of the WG parameter conditions. We also demonstrate how a mixture of various aggregated receptor-states in the multivalent models can lead to the transition of cooperativity from positive to negative.

\textit{Theoretical framework.} 
We consider the following assumptions: (a) there is no internalization of ligands and receptors; (b) each binding process is independent of chain and ring formation; (c) the four local equilibrium constants ($K_{x0}$,$K_{0}$,$K_{1}$,$K_{2}$) are dependent on each other, and; (d) detailed balance conditions are given by $K_{x1} = K_{x0} K_{1}/K_{0}$ and $K_{x2} = K_{x0} K_{1} K_{2}/K_{0}^2$.
%\begin{equation}
%K_{x1} = \frac{K_{x0} K_{1}}{K_{0}}\ \text{text}\ K_{x2} = \frac{K_{x0} K_{1} K_{2}}{K_{0}^2}\ .
%\end{equation}

In state-vector representation of physical observables, the dimerization model is described by a function containing the probabilities of biochemical interactions that form various aggregated receptor-states. All possible aggregated receptor-states in the dimerization models can be treated mathematically as basis vectors in a multidimensional real vector space. Observed receptor-state vectors in the formation of nulls (${\bf \Phi}$, ${\bf \Phi'}$), monomers (${\bf M}$, ${\bf M'}$) and dimers (${\bf D'}$) are given by
\begin{equation}
{\bf \Phi} = \left(\begin{array}{c}r \\rr \\rrr \\\vdots \\r^{^{N}} \end{array}\right)
\ \ 
{\bf M} = \left(\begin{array}{c}R \\Rr \\Rrr \\\vdots \\Rr^{^{N-1}} \end{array}\right)
\ \ 
{\bf \Phi'} = \left(\begin{array}{c}r\cdot r \\r\cdot rr \\r\cdot rrr \\\vdots \\r^{^{N}}\cdot r^{^{N}} \end{array}\right)
\nonumber
\end{equation}
\begin{equation}
{\bf M'} = \left(\begin{array}{c}R\cdot r \\R\cdot rr \\R\cdot rrr \\\vdots \\Rr^{^{N-1}}\cdot r^{^{N}} \end{array}\right)
\ \ 
{\bf D'} = \left(\begin{array}{c}R\cdot R \\R\cdot Rr \\R\cdot Rrr \\\vdots \\Rr^{^{N-1}}\cdot Rr^{^{N-1}} \end{array}\right)
\end{equation}
where $r$ and $R$ represent the receptors and the ligand-mediated receptors, respectively. $N$ refers the number of receptors that can be aggregated in the ${\bf \Phi}$ and ${\bf M}$ observed states. There are $N^2$ elements in the ${\bf \Phi'}$, ${\bf M'}$, and ${\bf D'}$ observed states. 

%\LARGE
%\begin{tabular}{ccccc}
%${\bf \Phi}$ & + & ${\bf \Phi}$ & $\bf\underset{d_{\scriptscriptstyle x0}}{\stackrel{k_{\scriptscriptstyle x0}}{\rightleftharpoons}}$ & ${\bf \Phi'}$ \\
%$\bf\ {\scriptscriptstyle d_{0}} \upharpoonleft\downharpoonright {\scriptscriptstyle Lk_{0}}$ & & & & $\bf\ {\scriptscriptstyle d_{1}} \upharpoonleft \downharpoonright {\scriptscriptstyle Lk_{1}}$ \\
%${\bf M}$ & + & ${\bf \Phi}$ & $\bf\underset{d_{\scriptscriptstyle x1}}{\stackrel{k_{\scriptscriptstyle x1}}{\rightleftharpoons}}$ & ${\bf M'}$ \\
% & & $\bf\ {\scriptscriptstyle d_{0}} \upharpoonleft\downharpoonright {\scriptscriptstyle Lk_{0}}$ & & $\bf\ {\scriptscriptstyle d_{2}} \upharpoonleft\downharpoonright {\scriptscriptstyle Lk_{2}}$ \\
%${\bf M}$ & + & ${\bf M}$ & $\bf\underset{d_{\scriptscriptstyle x2}}{\stackrel{k_{\scriptscriptstyle x2}}{\rightleftharpoons}}$ & ${\bf D'}$ 
%\end{tabular}

Figure~\ref{fig01;network} shows a network of observed receptor-states in the dimerization model. In first-order interactions of ligands with receptors, the rates of association (${\bf k_{0}}$, ${\bf k_{1}}$, ${\bf k_{2}}$) and dissociation (${\bf d_{0}}$, ${\bf d_{1}}$, ${\bf d_{2}}$) are represented by $N \times N$ and/or $N^2 \times N^2$ diagonal matrices acting upon the basis vectors, transforming one aggregated state to one observed state. 

The dissociation rates (${\bf d_{x0}}$, ${\bf d_{x1}}$, ${\bf d_{x2}}$) for the receptor-receptor interactions are represented by $N^2 \times N^2$ diagonal matrices. Non-diagonal matrices of the association rates can, however, transform into a mixture of various aggregated states in one observed state. The non-diagonal matrices can be written in the form of 
\begin{equation}
{\bf k_{x0}} = k_{x0} {\bf F_{x0}},
\ \ 
{\bf k_{x1}} = k_{x1} {\bf F_{x1}},
\ \ 
{\bf k_{x2}} = k_{x2} {\bf F_{x2}}
\end{equation}
where ${k_{xi}}$ and ${\bf F_{xi}}$ are the receptor-receptor association rate and $N \times N$ scaling matrices of the $i$-th index, respectively. 
%While the eigenvectors associated with these non-diagonal matrices represent the possible aggregated receptor-states in one observed state, the eigenvalues of the coupling matrices are real numbers that link to a measurable receptor-receptor coupling of the observed monomer and dimer formations in the aggregation system. 
%In our framework, a measured value of the receptor-receptor coupling is defined to be degenerate if there are at least two linearly independent aggregated receptor-states associated with the same eigenvalue.

For convenience, we redefine the dimensionless lumped parameter that constrains the fraction of dimer formations in the absence of ligands. The lumped parameter can be rewritten in the matrix form of
\begin{equation}
{\bf k_{x}} = k_{x} {\bf F_{x0}}
\end{equation}
where $k_x$ is the dimensionless lumped parameter originally defined in the WG formulations~\cite{wofsy1992a, *wofsy1992b}.

%The equations are given by 
%\begin{eqnarray}
%\sum_{i=0} \frac{d }{dt} {\Phi_i} & = & \sum^{N}_{i,j = 0} \left( - k^{0}_{ij} L \Phi_{j} + d^{0}_{ij} M_{j} \right) - \sum^{N}_{i,j = 0} G_{x0,ij} \Phi_{i} \Phi_{j} + \sum^{N'}_{k = 0} H_{0,kk} \Phi'_{k} - \sum^{N}_{i,j = 0} G_{x1,ij} \Phi_{i} M_{j} + \sum^{N'}_{k = 0} H_{1,kk} M'_{k} \nonumber\\
%\sum_{i=0} \frac{d }{dt} {\Phi'_i} & = & \sum^{N'}_{i,j = 0} \left( - k^{1}_{ij} L \Phi'_{j} + d^{1}_{ij} M'_{j} \right) + \sum^{N}_{i,j = 0} G_{x0,ij} \Phi^{i} \Phi_{j} - \sum^{N'}_{k = 0}   H_{0,kk} \Phi'_{k} \nonumber\\
%\sum_{i=0} \frac{d }{dt} {M_i} & = & \sum^{N}_{i,j = 0} \left( + k^{0}_{ij} L \Phi_{j} - d^{0}_{ij} M_{j} \right) - {\bf M^{T} G_{x1} \Phi} + \sum^{N'}_{k = 0} H_{1,kk} M'_{k}  - {\bf M^{T} G_{x2} M} + \sum^{N'}_{k = 0} H_{2,kk} M'_{k} \nonumber\\
%\sum_{i=0} \frac{d }{dt} {M'_i} & = & \sum^{N'}_{i,j = 0} \left( + k^{1}_{ij} L \Phi'_{j} - d^{1}_{ij} M'_{j} - k^{2}_{ij} L M'_{j} + d^{2}_{ij} D'_{j} \right) - {\bf M^{T} G_{x1} \Phi} + \sum^{N'}_{k = 0} H_{1,kk} M'_{k} \nonumber\\
%\sum_{i=0} \frac{d }{dt} {D'_i} & = & \sum^{N'}_{i,j = 0} \left( + k^{2}_{ij} L M'_{j} - d^{2}_{ij} D'_{j} \right) + {\bf M^{T} G_{x2} M} - \sum^{N'}_{k = 0} H_{2,kk} D'_{k} \nonumber
%\end{eqnarray}
%where {\bf G} and {\bf H} are 2nd-order association and dissociation coupling matrices, respectively. {\it others} means 1st-order reaction and molecular diffusion terms. Diffusion-terms are omitted in those equations. 
%

\textit{Multivalent models.} 
We construct monovalent ($N=1$) and bivalent ($N=2$) cell-models of ligand-induced receptor dimerization. We then use the E-Cell platform~\cite{tomita1999, arjunan2010} to simulate the cell-models of biological fluctuation that arise from stochastic changes in the cell surface geometry, number of receptors, ligand binding, molecular states, and diffusion constants. These cell-models assume that the non-diffusive receptors are uniformly distributed on a flat cell-surface measuring $100\ {\rm \mu m}$ and $100\ {\rm \mu m}$ in the horizontal and vertical axes. We also assume that the total receptor concentration, binding affinity and dissociation rates for each interaction are given by $T = 4.977\ {\rm \#receptors/\mu m^{2}}$, $K_{0} = 1.00\ {\rm nM}$, $d_0 = 0.01\ {\rm sec^{-1}}$, $d_1 = 10^{-5}\ {\rm sec^{-1}}$, $d_2 = 1.00\ {\rm sec^{-1}}$, and $d_{x0} = d_{x1} = d_{x2} = 1.00\ {\rm sec^{-1}}$. The relation of the local equilibrium constants to the association and dissociation rates is also given by $K_{i} = d_{i}/k_{i}$ where $i = 0, 1, 2, x0, x1, x2$. In a concentration range of ligand stimuli from $10^{-4}$ to $100\ {\rm nM}$, we run model-simulations for a period of $100,000\ {\rm sec}$ to verify the complete convergence of receptor response to full equilibrium.

The scaling factor and matrices in the monovalent model are given by $k_x = T/K_{x0}$ and ${\bf F_{x0}} = {\bf F_{x1}} = {\bf F_{x2}} = \alpha$. In the bivalent model, we assume that the symmetric scaling matrices can be written in the form of
\begin{equation}
{\bf F_{x0}} = {\bf F_{x1}} = 
\left(
\begin{array}{cc}
\alpha & \gamma\sqrt{\alpha\beta} \\
\gamma\sqrt{\alpha\beta} & \beta
\end{array}
\right),
\ \ 
{\bf F_{x2}} = 
\left(
\begin{array}{cc}
\alpha & 0 \\
0 & 0
\end{array}
\right)
\end{equation}
where $\alpha, \beta$ and $\gamma$ are matrix elements. $\gamma$ must be less than unity to satisfy the positive definite condition. The second-order interactions forming oligomers (e.g., dimers and trimers) in the null and monomeric observables are given by
\begin{eqnarray}
& {\bf \Phi^{\dag}}\ {\bf k_{x0}}\ {\bf \Phi} \label{eqn;null_obs} &\\
& {\bf M^{\dag}}\ {\bf k_{x1}}\ {\bf \Phi} \label{eqn;mono_obs} &
\end{eqnarray}
where $\bf \Phi^\dag$ and $\bf M^{\dag}$ represent dual vectors of the null and monomeric observables, respectively. In these formulations, the second-order interaction of the null observables can exhibit null dimers, trimers and tetramers:~$r + r \rightarrow rr$, $r + rr \rightarrow rrr$ and $rr + rr \rightarrow rrrr$. Monomeric dimers, trimers and tetramers can be also formed through the second-order interactions between the null and monomeric observables:~$R + r \rightarrow Rr$, $rR + r \rightarrow rRr$, $R + rr \rightarrow Rrr$ and $rR + rr \rightarrow rRrr$. There is no dimeric trimers and tetramers defined in the bivalent model.

\textit{The WG formulation}~\cite{wofsy1992a, *wofsy1992b}. 
The network diagram of the WG dimerization model is equivalent to that of the monovalent model. The dimerization process was, however, formulated under a ``special" assumption that the local equilibrium constant of direct receptor-receptor (or second-order) interaction ($K_{x0}$) is independent of first-order interactions of receptors associated with ligands ($K_{0}$,$K_{1}$,$K_{2}$). The total cell-surface receptor concentrations and the number of ligand-induced oligomers per a unit surface-area are given by
\begin{eqnarray}
T & = & \left( 1 + L K_0 \right) X + K_{x0} \left(1+2 L K_{1} + L^2 K_{1} K_{2} \right) X^2\hspace{0.4cm} \\
B & = & L K_{0} X + K_{x0} \left( L K_{1} + \frac{1}{2} L^2 K_{1}K_{2} \right) X^2
\label{eqn;bound}
\end{eqnarray}
where $X$ and $L$ are the concentration of unbound receptors and ligand concentration input in ${\rm nM}$, respectively. The $1/2$-factor in the third term of Eq.~(\ref{eqn;bound}) is meant to count the dimers as single molecules. Unit representations of local equilibrium constants in this formulation are not consistent with the units in our multivalent formulation, thereby requiring a unit transformation: $K_{i} \rightarrow K^{-1}_{i}$ where $i = 0, 1, 2, x0$. 

This WG formulation leads to a parameter condition that approximately exhibits negative cooperativity. The condition can be written in the form of 
\begin{eqnarray}
\frac{K_{1} \left( K_{1} - K_{2} \right)}{\left( K_{1} - K_{0} \right)^2} & \ge & \frac{\sqrt{1 + 4 k_{x}} - 1}{2 k_{x} \sqrt{1 + 4 k_{x}}}
\label{eqn;wg_condition}
\end{eqnarray}
where $k_x$  ($= T K_{x0}$) is the dimensionless lumped parameter. This relation implies that the model always exhibits positive cooperativity if $K_1 = K_2$.

%the local equilibrium constants for the first- and second-kind of ligand-receptor interactions are identical ($K_1 = K_2$). 
%Also, the WG model has been extended to include the formation of higher-order oligomers of unobserved receptors. The bivalent form of the extended WG model exhibits positive cooperativity if $K_1 = K_2$ (see appendix section in Wofsy et al. \cite{wofsy1992a}).

\begin{figure}
\leftline{\bf \hspace{0.04\linewidth} (a) \hspace{0.41\linewidth} (b)}
\centering
     \includegraphics[width=0.48\linewidth]{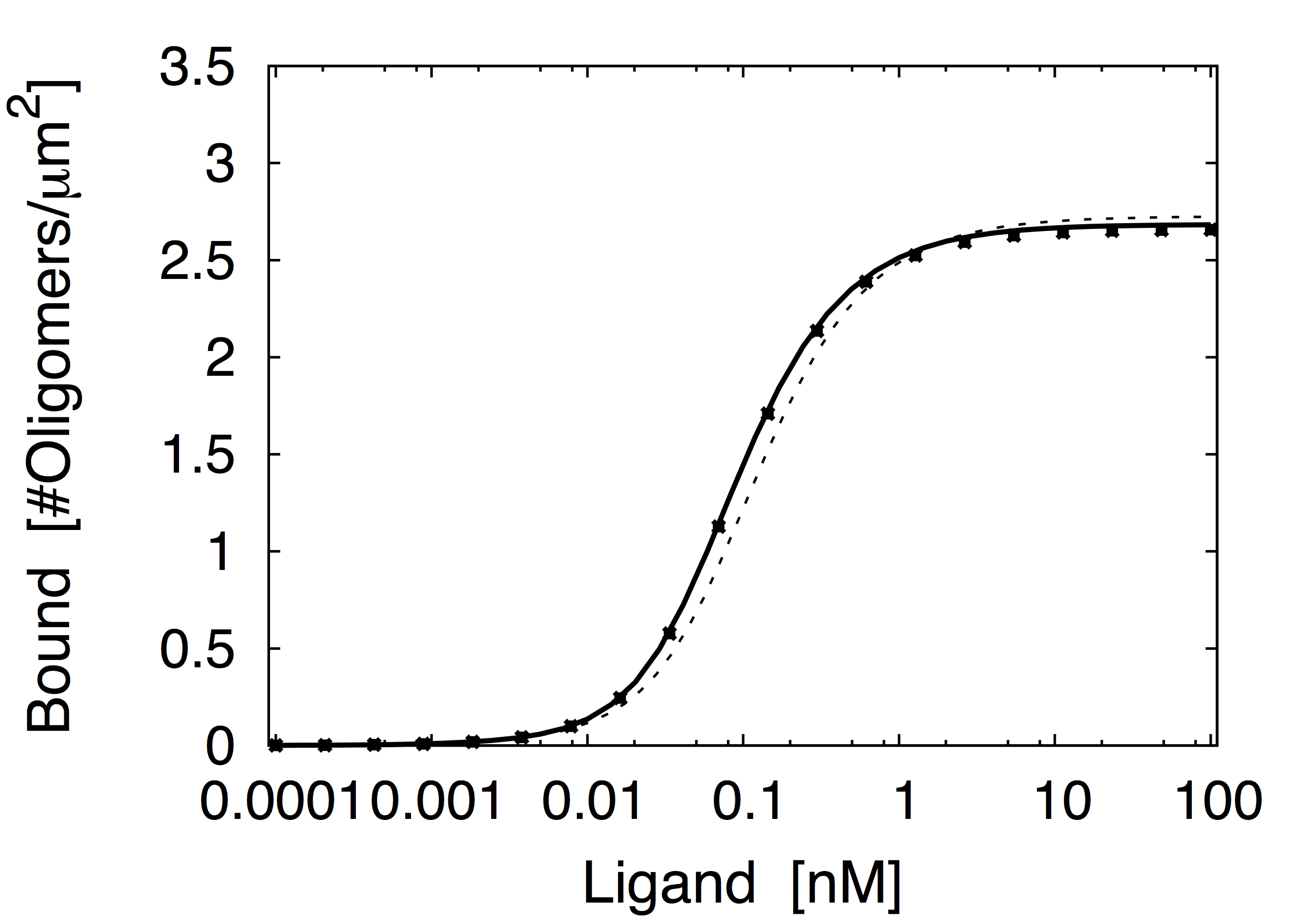}
     \includegraphics[width=0.48\linewidth]{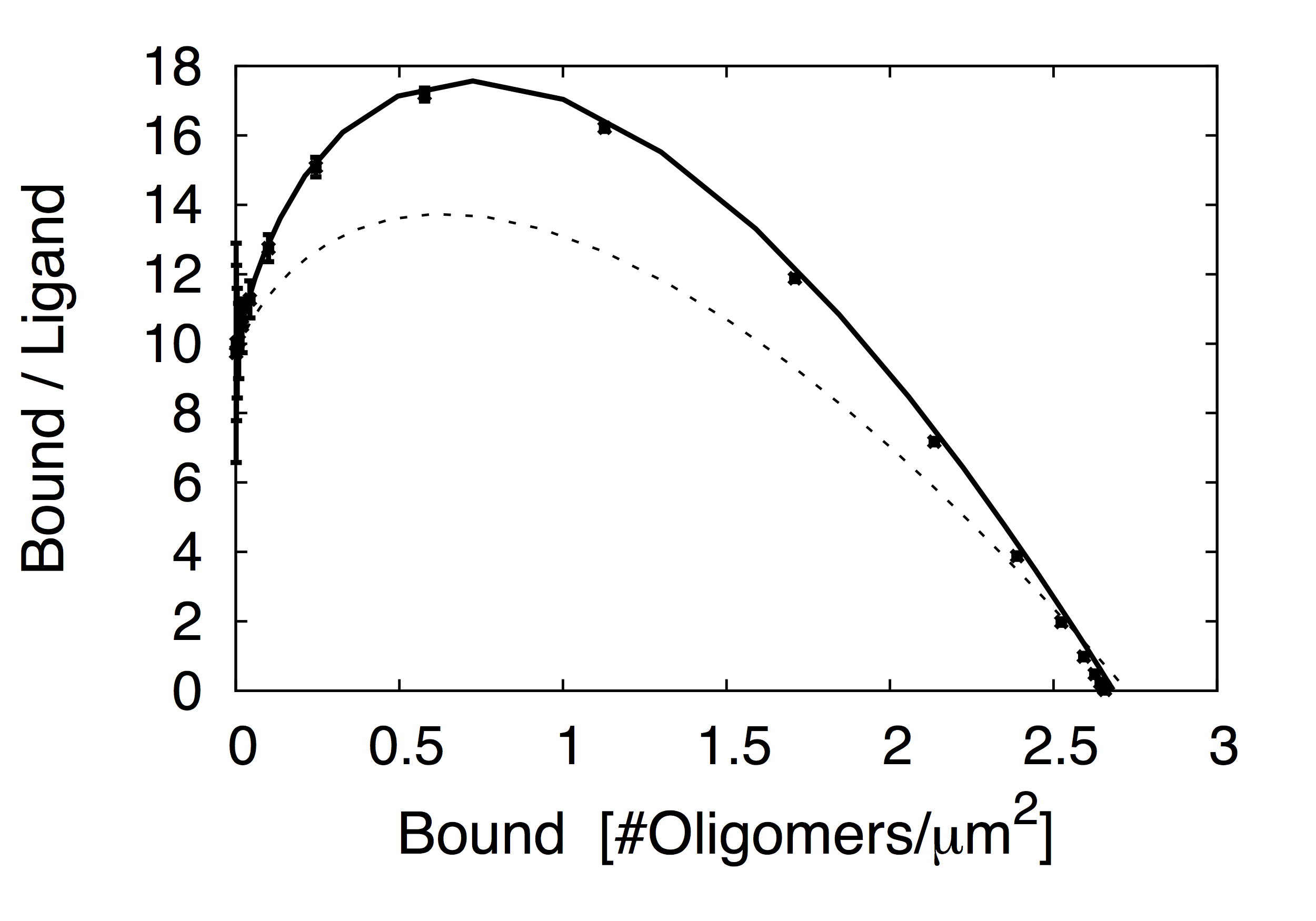}

\leftline{\bf \hspace{0.04\linewidth} (c) \hspace{0.41\linewidth} (d)}
\centering
     \includegraphics[width=0.48\linewidth]{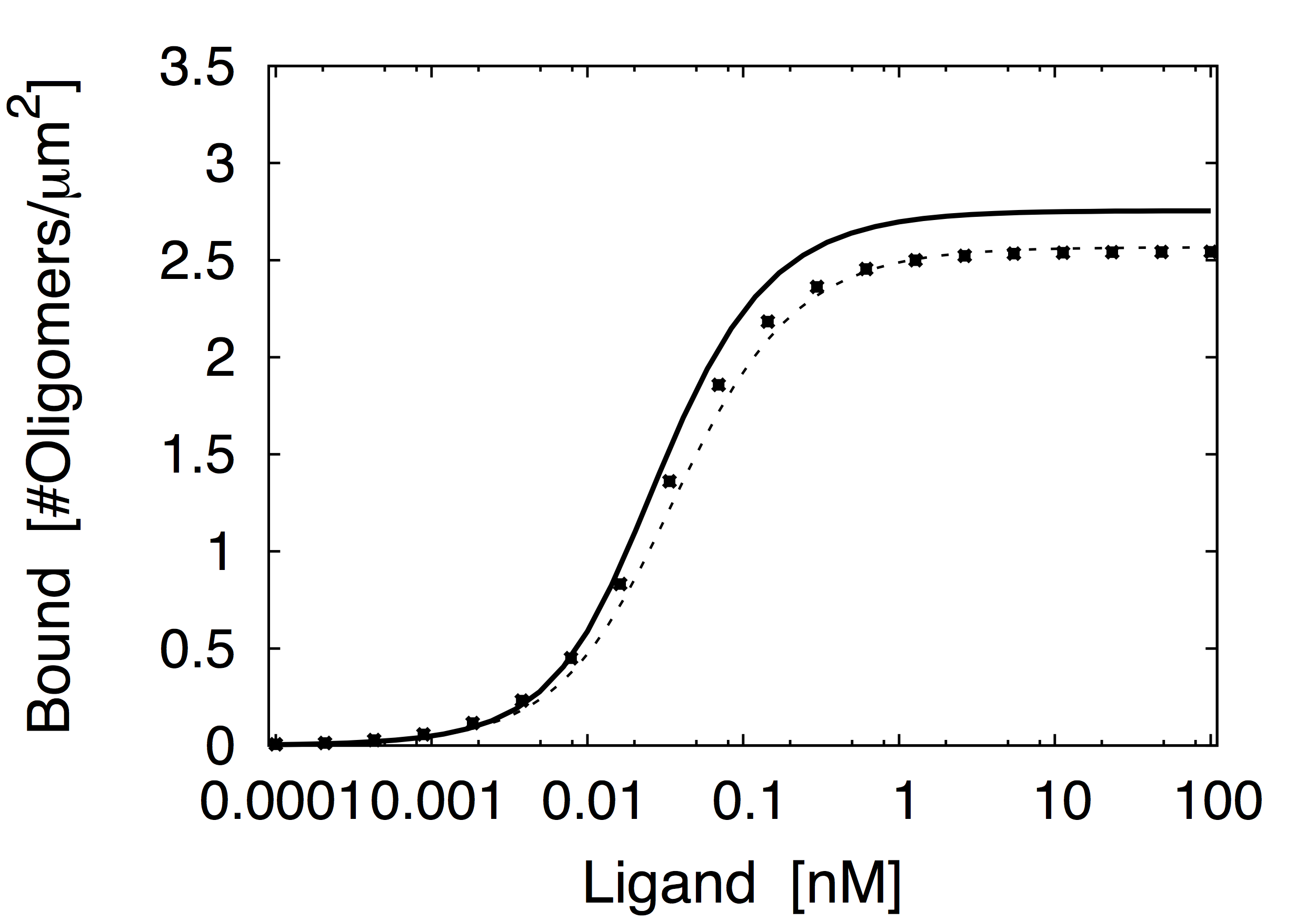}
     \includegraphics[width=0.48\linewidth]{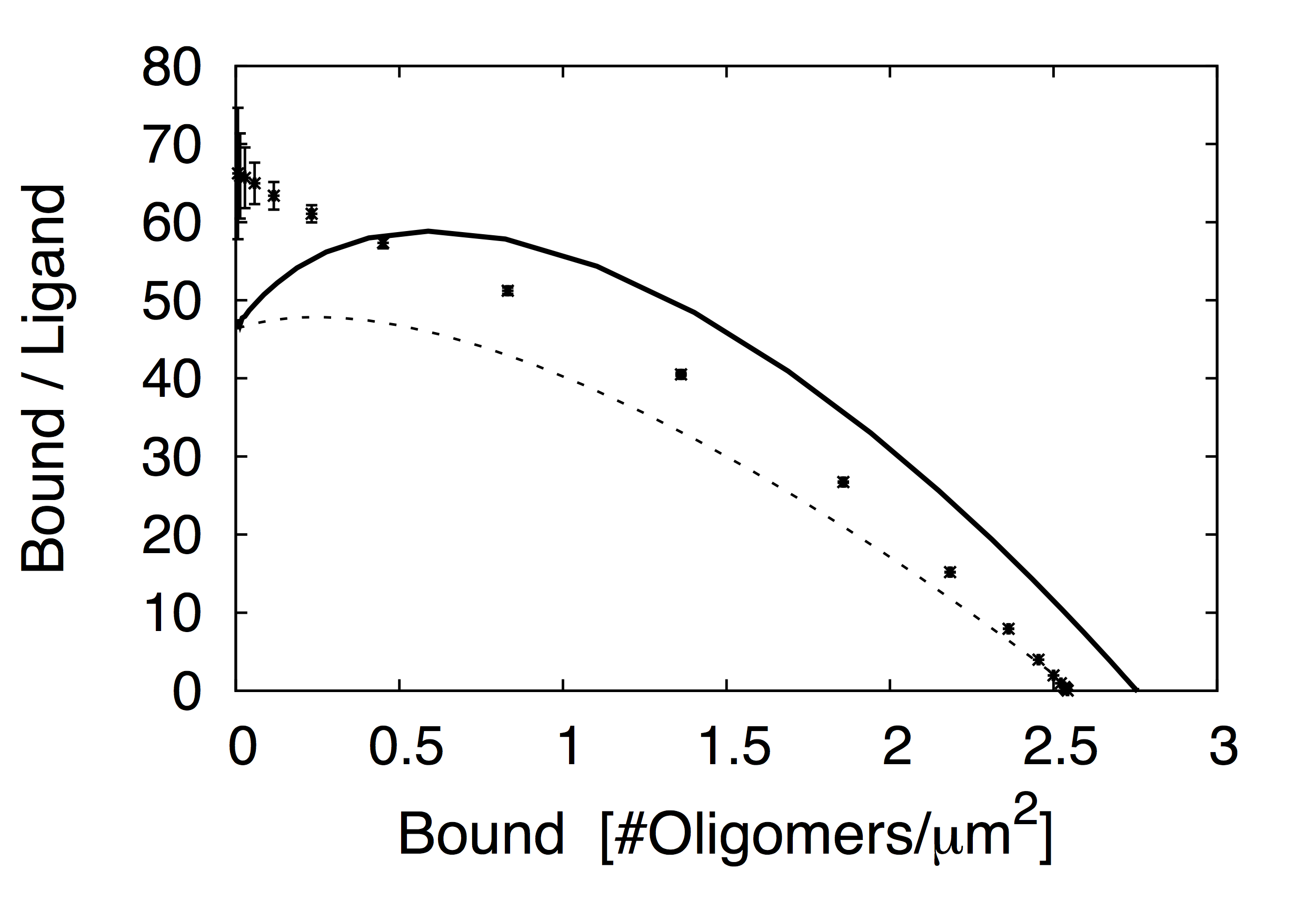}

\leftline{\bf \hspace{0.04\linewidth} (e) \hspace{0.41\linewidth} (f)}
\centering
     \includegraphics[width=0.48\linewidth]{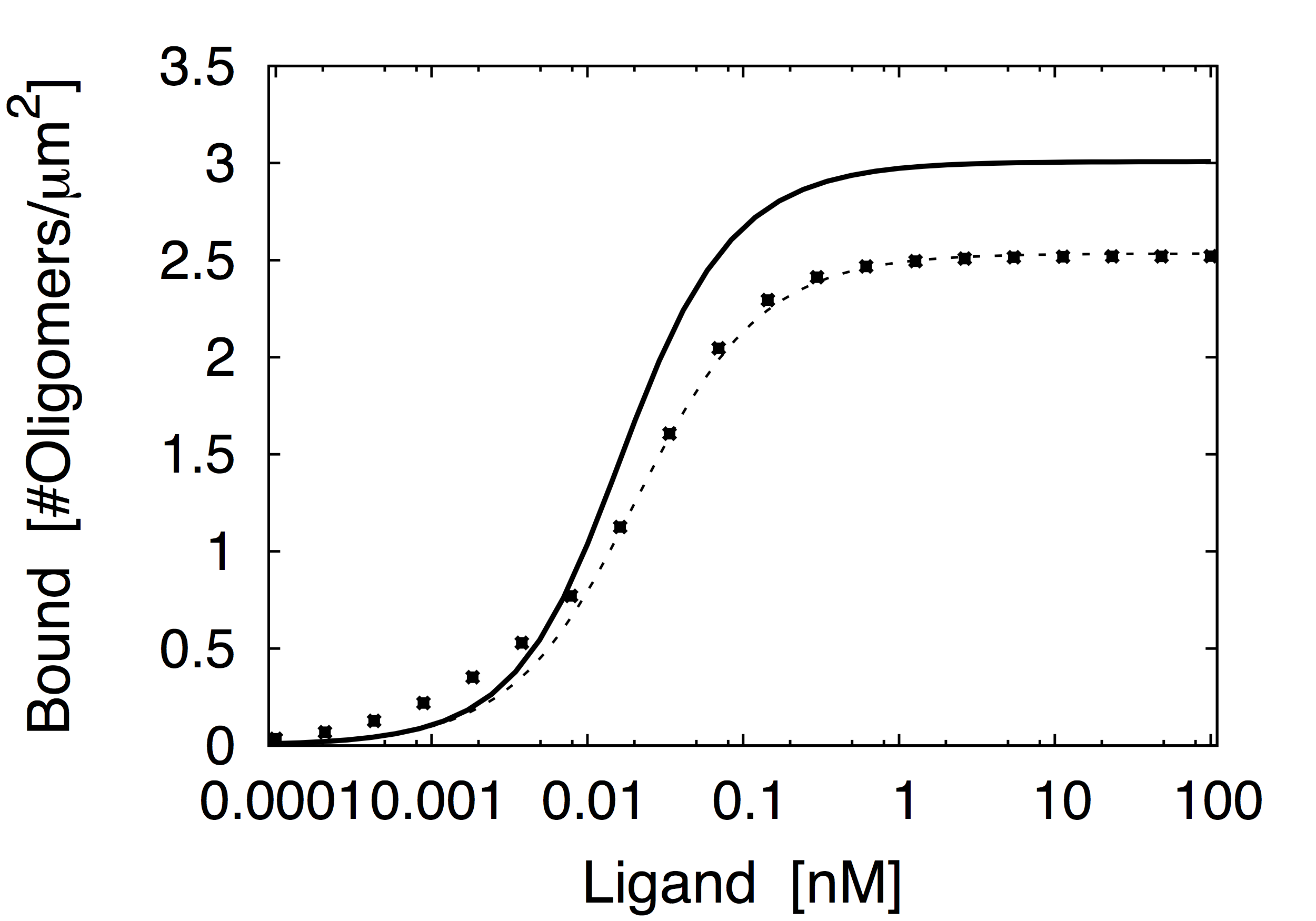}
     \includegraphics[width=0.48\linewidth]{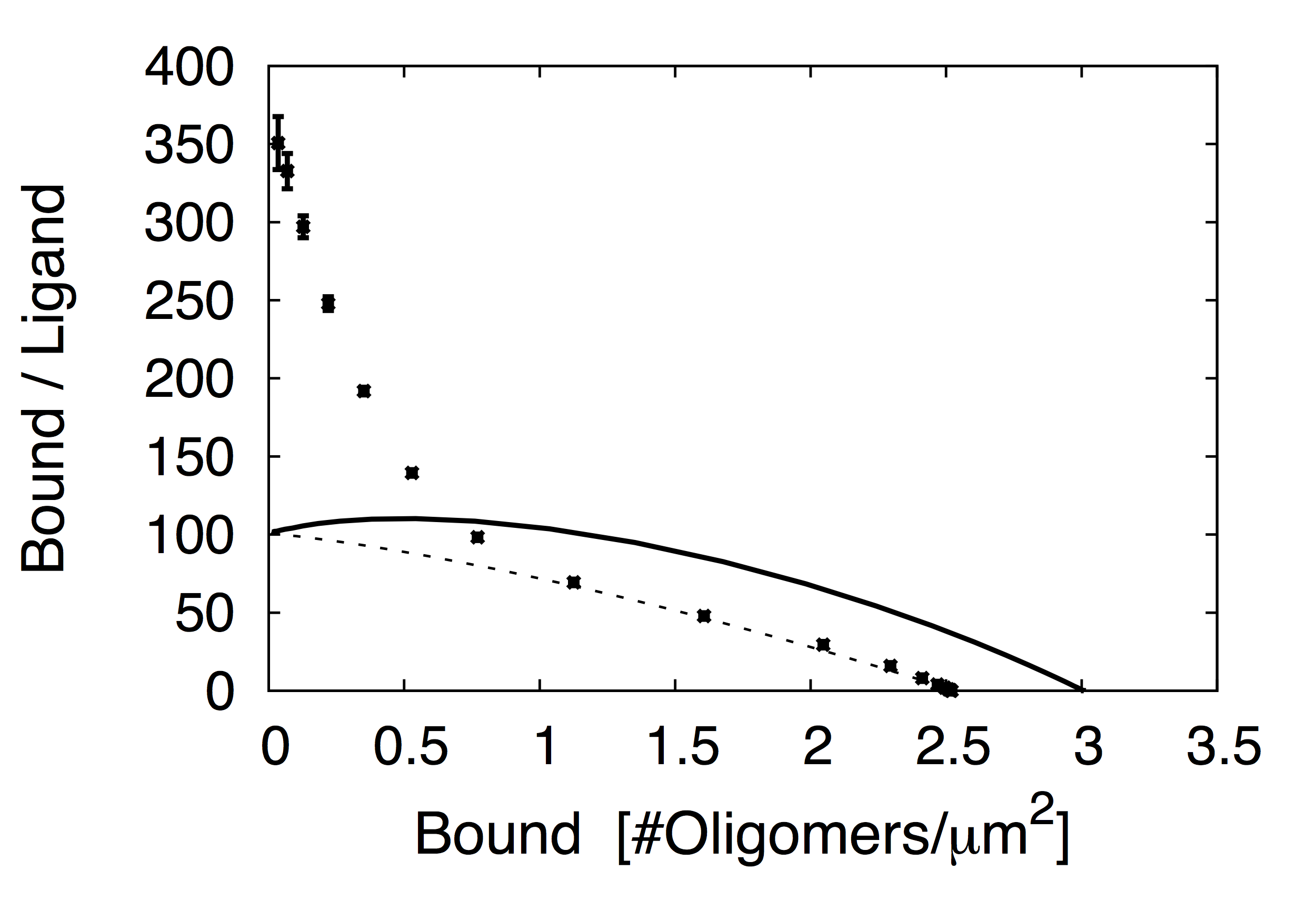}

\caption{{\bf Model comparison}. We compare the binding curves and Scatchard plots among the three models assuming $K_{1} = K_{2}= 100 K_{0}$: WG formulation, monovalent model ($\alpha = 1$), and bivalent model ($\alpha = \beta = 1$, $\gamma = 0$), for $k_x = 0.01$ (a,b), $k_x = 0.10$ (c,d), $k_x = 0.30$ (e,f). The solid and dashed black lines represent the response curves for the monovalent model and the WG formulation, respectively. Black crosses represent the bivalent model.}
\label{fig02;comparison} 
\end{figure}

\begin{figure*}
\leftline{\bf \hspace{0.03\linewidth} (a) \hspace{0.30\linewidth} (b) $\lambda_{+} = \lambda_{-}$ ($\gamma = \Delta = 0.00$) \hspace{0.08\linewidth} (c) $\lambda_{+} \neq \lambda_{-}$ ($\gamma = 0.10$)}
\centering
     \includegraphics[width=0.32\linewidth]{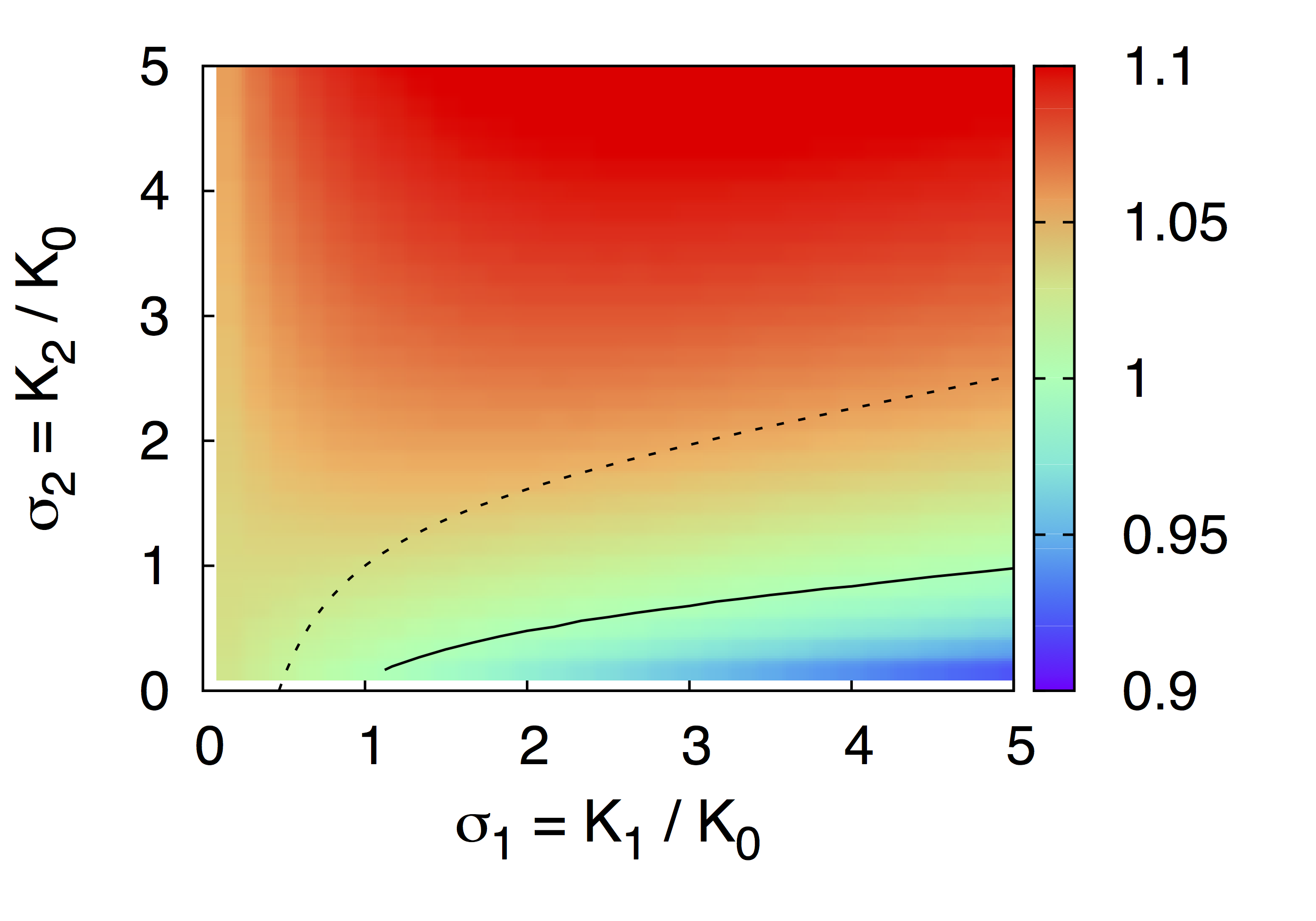}
     \includegraphics[width=0.32\linewidth]{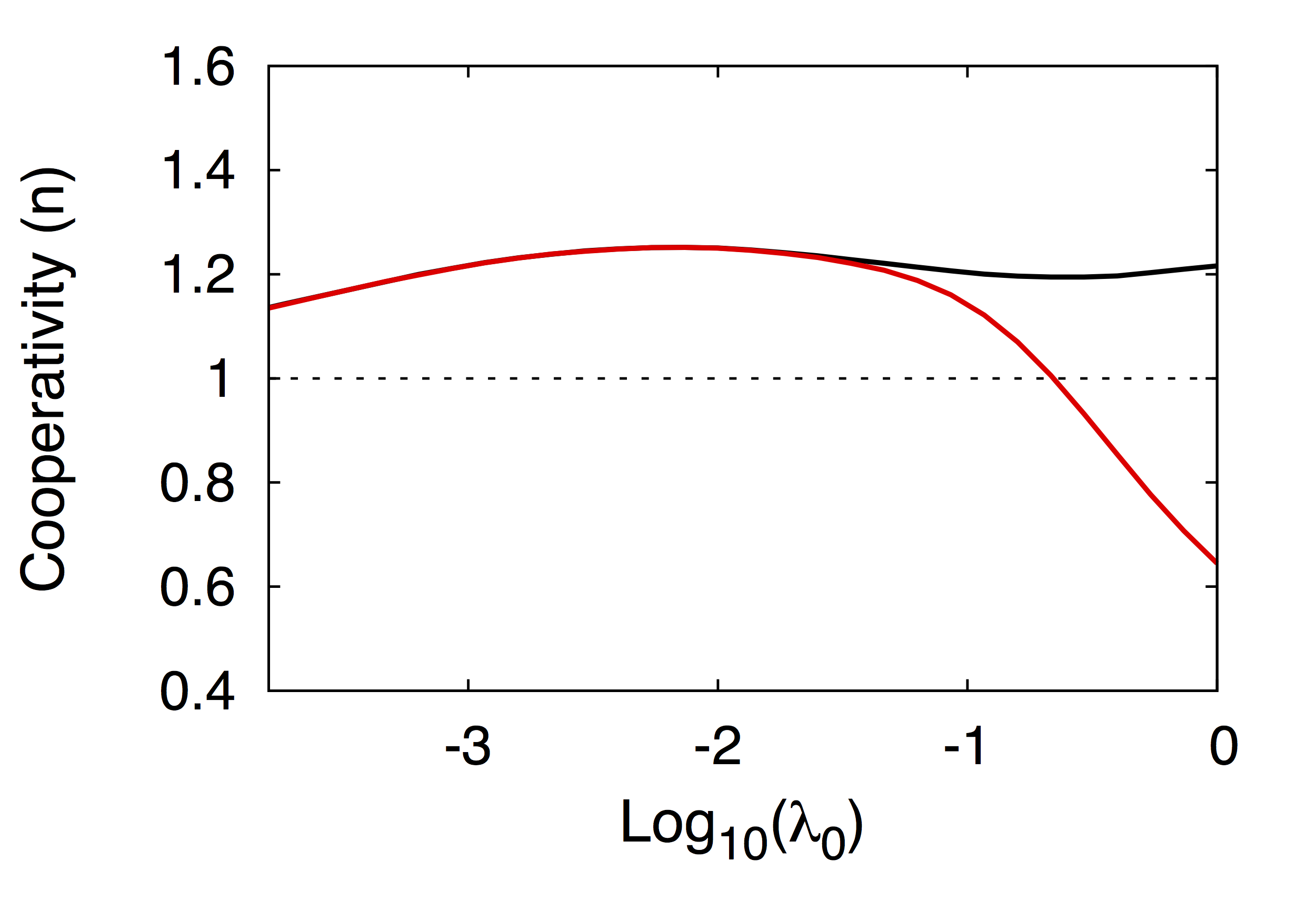}
     \includegraphics[width=0.32\linewidth]{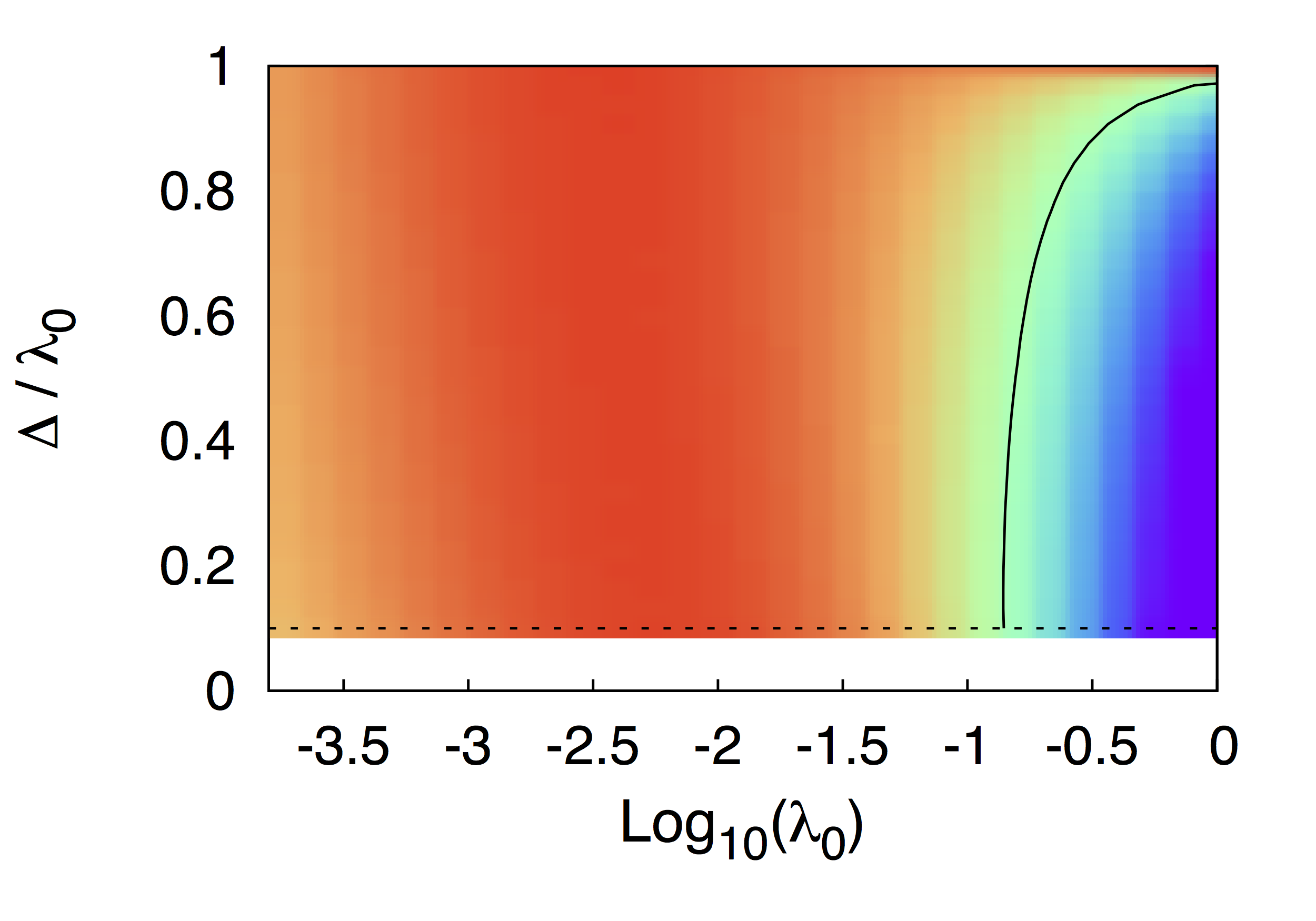}

\caption{{\bf Cooperativity of monovalent and bivalent models}. (a) Cooperativity of the monovalent model is shown as a function of the first and second ligand-receptor equilibrium constants, assuming $k_x = 0.10$. Colors represent cooperativity ($n$) of the monovalent model. The black solid and dashed lines represent no cooperativity ($n=1$) in the monovalent model and the WG condition given by Eq.~(\ref{eqn;wg_condition}), respectively. (b) and (c) Cooperativity of the bivalent model is shown as a function of the eigenvalues, assuming $K_{1} = K_{2}= 100 K_{0}$ and $k_x = 0.01$. $\alpha$ is varied from $0.01$ to $100$. (b) $\lambda_{+} = \lambda_{-}$ ($\gamma = \Delta = 0.00$). The black and red solid lines represent cooperativity of the monovalent and bivalent models as a function of $\lambda_0$, respectively. The black dashed line corresponds to no cooperativity ($n = 1$). (c) $\lambda_{+} \neq \lambda_{-}$ ($\gamma = 0.10$). The black line represents no cooperativity. Colors represent cooperativity ($n$) in the range of $0.7$ (blue) to $1.3$ (red). The black dashed line is the physical limit to satisfy the positive definite condition.}
\label{fig03;coop}
\end{figure*}

\textit{Model comparison.}
Cooperative effects can be generally seen in the concavity of the Scatchard plots. We compare the Scatchard plots among the three cell-models configured to have same parameter values. Model differences can be clearly seen in Figure~\ref{fig02;comparison}b,d,f. For $k_x = 0.01$ and $0.10$, the Scatchard plots of the three models exhibit concave downward curvatures that approximately represent positive cooperativity. The bivalent model ($k_x = 0.30$) in Figure~\ref{fig02;comparison}f, however, exhibits a concave-up curve of the Scatchard plot that represents negative cooperativity. If the Scatchard plot represents a straight line, then there is no cooperativity. These comparison results thus imply that cooperative effects can vary as a function of the lumped parameter, $k_x$. 

Cooperative effects can be also seen in the shape of the equilibrium binding curves. We compare the binding curves among the three cell-models configured to have same parameter values. For $k_x = 0.01$, $0.10$ and $0.30$, the comparison results are shown in Figure~\ref{fig02;comparison}a,c,e. In standard systems biology approaches, the Hill function can be fitted to the binding curves to quantify the cooperative characteristics in the cell-models. The Hill function can be generally written in the form of
\begin{eqnarray}
B(L) & = & \frac{B_0 L^n}{K^n_A + L^n}
\label{eqn;hill}
\end{eqnarray}
where $L$, $B_0$, $K_A$ and $n$ represent ligand concentration, maximum area-density of the ligand-induced oligomers, ligands occupying half of the oligomers and the Hill coefficient, respectively. If the Hill coefficient is less than unity ($n<1$), then the receptor system exhibits negative cooperativity. If $n>1$, then cooperativity is positive. There is no cooperativity if $n=1$. 

%\begin{figure}
%\centering
%     \includegraphics[width=1.00\linewidth]{figures/figure_03.png}
%
%\caption{{\bf Cooperativity of monovalent model in $\sigma_1$-$\sigma_2$ space}. Cooperativity of the monovalent model is shown as a function of the first and second ligand-receptor equilibrium constants, assuming $k_x = 0.10$. Colors represent cooperativity ($n$) of the monovalent model. The black solid and dashed lines represent no cooperativity ($n=1$) in the monovalent model and the WG condition given by Eq.~(\ref{eqn;wg_condition}), respectively.}
%\label{fig03;coop_s1s2}
%\end{figure}

For a fixed $k_x = 0.10$, the best fit Hill coefficients of the monovalent model are mapped as a function of $K_1$ and $K_2$. In Figure~\ref{fig03;coop}a the cooperativity mapping result is shown and compared with the WG condition given by Eq.~(\ref{eqn;wg_condition}). This comparison result clearly shows suppression of the negative cooperative region in the monovalent model, implying inconsistent cooperative responses between the monovalent model and the WG formulation.

\textit{Diagonalization.} 
To see the cooperative effects that arise from the second-order interactions forming the higher-order oligomers in the bivalent model [see Eq.~(\ref{eqn;null_obs}) and (\ref{eqn;mono_obs})], we diagonalize the lumped parameter matrix ${\bf k_{x}}$. Eigenvalues are given by
\begin{eqnarray}
\lambda_{\pm} & = & \frac{k_{x}}{2} \left[ \left( \alpha + \beta \right) \pm \sqrt{\left( \alpha - \beta \right)^2 + 4 \gamma^2\alpha\beta}\ \right] \nonumber \\
& = & \lambda_0 \pm \Delta
\label{eqn;ev}
\end{eqnarray}
where the dynamic range is $\gamma\lambda_0 \le \Delta \le \lambda_0$. 
%If there are two linearly independent eigenvectors having the same eigenvalue ($\lambda_{+} = \lambda_{-}$), then the eigenvalues of the two eigenvector are degenerate. The eigenvalues are non-degenerate if $\lambda_{+} \neq \lambda_{-}$. 

We compare cooperative responses between the monovalent and bivalent models when the eigenvalues of the lumped parameter matrix in the bivalent model are identical ($\lambda_{+} = \lambda_{-}$). Figure~\ref{fig03;coop}b shows cooperativity transition of the bivalent model as a function of the $\lambda_0$ component in Eq.~(\ref{eqn;ev}). While cooperativity is always positive in the monovalent model (black line), cooperativity in the bivalent model is shifted from positive to negative through the increase of $\lambda_0$ (red line). In the lower $\lambda_0$-range, the second-order couplings are weakly linked with cooperative responses, displaying identical cooperativity between the monovalent and bivalent models. The higher $\lambda_0$-values in the bivalent model can, however, increase the number of trimers ($rrR$) and tetramers ($rrrR$) in the monomeric observable $\bf M'$. Also, these higher-order oligomers are weakly linked with the fist-order couplings that represent ligand-dissociation ($rrR \rightarrow rrr$, $rrrR \rightarrow rrrr$) and -association ($rrR \rightarrow rRR$, $rrrR \rightarrow rrRR$). Because of these model parameter relations, the bivalent model exhibits the transition of cooperativity in the higher $\lambda_0$-range.

%The higher $\lambda_0$-values in the bivalent model can, however, increase the number of the trimers and tetramers in the monomeric observable $\bf M'$; also, these higher-order oligomers are weakly coupled with the fist-order interactions that represent ligand dissociation and association ($\bf M' \rightarrow  \Phi'$ and $\bf M' \rightarrow D'$). These parameter relations thus lead to the transition of cooperativity in the higher $\lambda_0$-range.
%
%We compare the cooperative responses between the monovalent and bivalent models if the eigenvalues of the lumped parameter matrix in the bivalent model are degenerate: two measurable receptor-receptor couplings are identical and indistinguishable. 

We also evaluate cooperative responses in the bivalent model in the case of differing eigenvalues ($\lambda_{+} \neq \lambda_{-}$). Figure~\ref{fig03;coop}c shows cooperativity of the bivalent model as a function of the $\lambda_0$ and $\Delta$ components in Eq.~(\ref{eqn;ev}).  As the ratio of these components converges to unity $\Delta/\lambda_0 \rightarrow 1$ (or $\beta \rightarrow 0$), the bivalent model becomes equivalent to the monovalent model, exhibiting positive cooperativity in the full $\lambda_0$-range. While cooperativity is always positive at $\Delta/\lambda_0 = 1$ (or $\beta = 0$), the trimer and tetramer formations in the monomeric observable can change cooperativity from positive (red region) to negative (blue region) for $\Delta/\lambda_0 < 1$ (or $\beta > 0$).

Our results may have biophysical implications in data-driven modeling approaches. Many biomolecular binding processes have been modeled under the scenario that the second-order couplings are weakly linked with cooperative effects. For example, the dimerization model that describes heregulin (HRG) binding with ErbB receptors in living cells was constructed using fluorescence microscopy images via HRG tagged with tetramethylrhodamine~\cite{hiroshima2012}. In this model, the first-order coupling representing a conformational change of receptor complex in the monomeric observable ($r R \leftrightarrow r' R$) plays a key role that gives rise to negative cooperativity. Nevertheless, there exist various aggregated receptor-states hidden in the physical observables. These states are realistic but unobserved through the fluorescent imaging techniques. By incorporating the second-order interactions forming higher-order oligomers that can include unobserved receptors into the dimerization model, cooperativity may shift from positive to negative, and vice versa. Because of these transitions, the data-driven modeling approach is not a straightforward method to identify physical sources giving rise to negative cooperative effects in the receptor systems.

\textit{Conclusion}. 
In this article, we explored the origin of negative cooperativity in dimer formations for equilibrium binding of ligands to cell-surface receptors, in terms of biochemical parameters for associations and dissociations. While receptor-receptor couplings in the ligand-induced receptor dimerization have been previously linked with cooperativity via minimal representations of physical observables, cooperative effects that arise from the mixture of various aggregated receptor-states hidden in each observed receptor-states have received less attention. In particular, we examined the cooperativity of monovalent and bivalent models. Our results from model simulations showed the suppression of negative cooperative regions in the monovalent model, thereby implying violation of parameter conditions expected from the WG formulation. We also demonstrated that the presence of higher-order oligomer formations in the bivalent model leads to the transition of cooperativity from positive to negative, thus affecting physical interpretations of the cooperativity extracted from data-driven modeling approaches. Furthermore, it is interesting to extend our state-vector representations to more general model frameworks by incorporating heterodimer formations of receptor-family members (e.g., dimerization of ErbB3 and EGF receptors~\cite{itano2018, ito2019}).

%Thus, by introducing hidden components (or states) in the network models, the cooperativity transitions can lead to various physical interpretations and meanings of the results extracted from the data-driven modeling approaches. 
%Such cooperativity shifts thus imply that physical interpretations and meanings of the results extracted from the data-driven modeling approaches can alter by introducing hidden components (or states) in the network models.
%Our work in this direction is underway.

%\begin{acknowledgments}
\textit{Acknowledgments.} 
We would like to thank Yasushi Okada, Tomonobu M. Watanabe, Jun Kozuka, Michio Hiroshima, Kozo Nishida, Suguru Kato, Toru Niina, Koji Ochiai, Keiko Itano, Kotone Itaya and Kaoru Ikegami for their guidance and support throughout this research work. We would also like to thank Kylius Wilkins for critical reading of the manuscript. This research work is supported by JSPS (Japanese Society for the Promotion of Science) KAKENHI Grant No. 15K12146.
%\end{acknowledgments}

%\nocite{apsrev42Control}
%\bibliographystyle{apsrev4-2}
%\bibliography{ref2019}% Produces the bibliography via BibTeX.

\begin{thebibliography}{25}%
\makeatletter
\providecommand \@ifxundefined [1]{%
 \@ifx{#1\undefined}
}%
\providecommand \@ifnum [1]{%
 \ifnum #1\expandafter \@firstoftwo
 \else \expandafter \@secondoftwo
 \fi
}%
\providecommand \@ifx [1]{%
 \ifx #1\expandafter \@firstoftwo
 \else \expandafter \@secondoftwo
 \fi
}%
\providecommand \natexlab [1]{#1}%
\providecommand \enquote  [1]{``#1''}%
\providecommand \bibnamefont  [1]{#1}%
\providecommand \bibfnamefont [1]{#1}%
\providecommand \citenamefont [1]{#1}%
\providecommand \href@noop [0]{\@secondoftwo}%
\providecommand \href [0]{\begingroup \@sanitize@url \@href}%
\providecommand \@href[1]{\@@startlink{#1}\@@href}%
\providecommand \@@href[1]{\endgroup#1\@@endlink}%
\providecommand \@sanitize@url [0]{\catcode `\\12\catcode `\$12\catcode
  `\&12\catcode `\#12\catcode `\^12\catcode `\_12\catcode `\%12\relax}%
\providecommand \@@startlink[1]{}%
\providecommand \@@endlink[0]{}%
\providecommand \url  [0]{\begingroup\@sanitize@url \@url }%
\providecommand \@url [1]{\endgroup\@href {#1}{\urlprefix }}%
\providecommand \urlprefix  [0]{URL }%
\providecommand \Eprint [0]{\href }%
\providecommand \doibase [0]{https://doi.org/}%
\providecommand \selectlanguage [0]{\@gobble}%
\providecommand \bibinfo  [0]{\@secondoftwo}%
\providecommand \bibfield  [0]{\@secondoftwo}%
\providecommand \translation [1]{[#1]}%
\providecommand \BibitemOpen [0]{}%
\providecommand \bibitemStop [0]{}%
\providecommand \bibitemNoStop [0]{.\EOS\space}%
\providecommand \EOS [0]{\spacefactor3000\relax}%
\providecommand \BibitemShut  [1]{\csname bibitem#1\endcsname}%
\let\auto@bib@innerbib\@empty
%</preamble>
\bibitem [{\citenamefont {Strandburg-Peshkin}\ \emph
  {et~al.}(2017)\citenamefont {Strandburg-Peshkin}, \citenamefont {Farine},
  \citenamefont {Crofoot},\ and\ \citenamefont {Couzin}}]{strandburg2017}%
  \BibitemOpen
  \bibfield  {author} {\bibinfo {author} {\bibfnamefont {A.}~\bibnamefont
  {Strandburg-Peshkin}}, \bibinfo {author} {\bibfnamefont {D.~R.}\ \bibnamefont
  {Farine}}, \bibinfo {author} {\bibfnamefont {M.~C.}\ \bibnamefont
  {Crofoot}},\ and\ \bibinfo {author} {\bibfnamefont {I.~D.}\ \bibnamefont
  {Couzin}},\ }\bibfield  {title} {\bibinfo {title} {{Habitat and social
  factors shape individual decisions and emergent group structure during baboon
  collective movement}},\ }\href {https://doi.org/10.7554/eLife.19505}
  {\bibfield  {journal} {\bibinfo  {journal} {eLife}\ }\textbf {\bibinfo
  {volume} {6}},\ \bibinfo {pages} {e19505} (\bibinfo {year}
  {2017})}\BibitemShut {NoStop}%
\bibitem [{\citenamefont {Strandburg-Peshkin}\ \emph
  {et~al.}(2015)\citenamefont {Strandburg-Peshkin}, \citenamefont {Farine},
  \citenamefont {Couzin},\ and\ \citenamefont {Crofoot}}]{strandburg2015}%
  \BibitemOpen
  \bibfield  {author} {\bibinfo {author} {\bibfnamefont {A.}~\bibnamefont
  {Strandburg-Peshkin}}, \bibinfo {author} {\bibfnamefont {D.~R.}\ \bibnamefont
  {Farine}}, \bibinfo {author} {\bibfnamefont {I.~D.}\ \bibnamefont {Couzin}},\
  and\ \bibinfo {author} {\bibfnamefont {M.~C.}\ \bibnamefont {Crofoot}},\
  }\bibfield  {title} {\bibinfo {title} {{Shared decision-making drives
  collective movement in wild baboons}},\ }\href@noop {} {\bibfield  {journal}
  {\bibinfo  {journal} {Science}\ }\textbf {\bibinfo {volume} {348}},\ \bibinfo
  {pages} {1358--1361} (\bibinfo {year} {2015})}\BibitemShut {NoStop}%
\bibitem [{\citenamefont {Koshland}\ and\ \citenamefont
  {Hamadani}(2002)}]{koshland2002}%
  \BibitemOpen
  \bibfield  {author} {\bibinfo {author} {\bibfnamefont {D.~E.}\ \bibnamefont
  {Koshland}}\ and\ \bibinfo {author} {\bibfnamefont {K.}~\bibnamefont
  {Hamadani}},\ }\bibfield  {title} {\bibinfo {title} {{Proteomics and models
  for enzyme cooperativity}},\ }\href {https://doi.org/10.1074/jbc.R200014200}
  {\bibfield  {journal} {\bibinfo  {journal} {J. Biol. Chem.}\ }\textbf
  {\bibinfo {volume} {277}},\ \bibinfo {pages} {46841--46844} (\bibinfo {year}
  {2002})}\BibitemShut {NoStop}%
\bibitem [{\citenamefont {Ferrell}(2009)}]{ferrell2009}%
  \BibitemOpen
  \bibfield  {author} {\bibinfo {author} {\bibfnamefont {J.~E.}\ \bibnamefont
  {Ferrell}},\ }\bibfield  {title} {\bibinfo {title} {{Q {\&} A :
  Cooperativity}},\ }\href@noop {} {\bibfield  {journal} {\bibinfo  {journal}
  {J. Biol.}\ }\textbf {\bibinfo {volume} {8}} (\bibinfo {year}
  {2009})}\BibitemShut {NoStop}%
\bibitem [{\citenamefont {Stefan}\ and\ \citenamefont
  {Nove}(2013)}]{stefan2013}%
  \BibitemOpen
  \bibfield  {author} {\bibinfo {author} {\bibfnamefont {M.~I.}\ \bibnamefont
  {Stefan}}\ and\ \bibinfo {author} {\bibfnamefont {N.~L.}\ \bibnamefont
  {Nove}},\ }\bibfield  {title} {\bibinfo {title} {{Cooperative Binding}},\
  }\href {https://doi.org/10.1371/journal.pcbi.1003106} {\bibfield  {journal}
  {\bibinfo  {journal} {PLOS Comput. Biol.}\ }\textbf {\bibinfo {volume} {9}},\
  \bibinfo {pages} {e1003106} (\bibinfo {year} {2013})}\BibitemShut {NoStop}%
\bibitem [{\citenamefont {Phillips}\ \emph {et~al.}(2013)\citenamefont
  {Phillips}, \citenamefont {Kondev}, \citenamefont {Theriot}, \citenamefont
  {Garcia},\ and\ \citenamefont {Orme}}]{phillips2013}%
  \BibitemOpen
  \bibfield  {author} {\bibinfo {author} {\bibfnamefont {R.}~\bibnamefont
  {Phillips}}, \bibinfo {author} {\bibfnamefont {J.}~\bibnamefont {Kondev}},
  \bibinfo {author} {\bibfnamefont {J.}~\bibnamefont {Theriot}}, \bibinfo
  {author} {\bibfnamefont {H.~G.}\ \bibnamefont {Garcia}},\ and\ \bibinfo
  {author} {\bibnamefont {Orme}},\ }\href@noop {} {\emph {\bibinfo {title}
  {Physical Biology of the Cell.}}},\ \bibinfo {edition} {2nd}\ ed.\ (\bibinfo
  {publisher} {Garland Science},\ \bibinfo {year} {2013})\BibitemShut {NoStop}%
\bibitem [{\citenamefont {Monod}\ \emph {et~al.}(1965)\citenamefont {Monod},
  \citenamefont {Wyman},\ and\ \citenamefont {Changeux}}]{monod1965}%
  \BibitemOpen
  \bibfield  {author} {\bibinfo {author} {\bibfnamefont {J.}~\bibnamefont
  {Monod}}, \bibinfo {author} {\bibfnamefont {J.}~\bibnamefont {Wyman}},\ and\
  \bibinfo {author} {\bibfnamefont {J.~P.}\ \bibnamefont {Changeux}},\
  }\bibfield  {title} {\bibinfo {title} {{On the nature of allosteric
  transitions: a plausible model.}},\ }\href
  {https://doi.org/10.1021/bi00865a047} {\bibfield  {journal} {\bibinfo
  {journal} {J. Mol. Biol.}\ }\textbf {\bibinfo {volume} {12}},\ \bibinfo
  {pages} {88--118} (\bibinfo {year} {1965})}\BibitemShut {NoStop}%
\bibitem [{\citenamefont {Koshland}\ \emph {et~al.}(1966)\citenamefont
  {Koshland}, \citenamefont {Nemethy},\ and\ \citenamefont
  {Filmer}}]{koshland1966}%
  \BibitemOpen
  \bibfield  {author} {\bibinfo {author} {\bibfnamefont {D.~E.~J.}\
  \bibnamefont {Koshland}}, \bibinfo {author} {\bibfnamefont {D.}~\bibnamefont
  {Nemethy}},\ and\ \bibinfo {author} {\bibfnamefont {G.}~\bibnamefont
  {Filmer}},\ }\bibfield  {title} {\bibinfo {title} {{Comparison of
  Experimental Binding Data and Theoretical Models in Proteins Containing
  Subunits}},\ }\href {https://doi.org/10.1021/bi00865a047} {\bibfield
  {journal} {\bibinfo  {journal} {Biochemistry}\ }\textbf {\bibinfo {volume}
  {5}},\ \bibinfo {pages} {365--385} (\bibinfo {year} {1966})}\BibitemShut
  {NoStop}%
\bibitem [{\citenamefont {Alan}(1999)}]{alan1999}%
  \BibitemOpen
  \bibfield  {author} {\bibinfo {author} {\bibfnamefont {F.}~\bibnamefont
  {Alan}},\ }\href@noop {} {\emph {\bibinfo {title} {Structure and mechanism in
  protein science : a guide to enzyme catalysis and protein folding.}}},\
  \bibinfo {edition} {1st}\ ed.\ (\bibinfo  {publisher} {W. H. Freeman},\
  \bibinfo {year} {1999})\BibitemShut {NoStop}%
\bibitem [{\citenamefont {Wofsy}\ \emph {et~al.}(1992)\citenamefont {Wofsy},
  \citenamefont {Goldstein}, \citenamefont {Lund},\ and\ \citenamefont
  {Wiley}}]{wofsy1992a}%
  \BibitemOpen
  \bibfield  {author} {\bibinfo {author} {\bibfnamefont {C.}~\bibnamefont
  {Wofsy}}, \bibinfo {author} {\bibfnamefont {B.}~\bibnamefont {Goldstein}},
  \bibinfo {author} {\bibfnamefont {K.}~\bibnamefont {Lund}},\ and\ \bibinfo
  {author} {\bibfnamefont {H.~S.}\ \bibnamefont {Wiley}},\ }\bibfield  {title}
  {\bibinfo {title} {{Implications of epidermal growth factor (EGF) induced egf
  receptor aggregation.}},\ }\href
  {https://doi.org/10.1016/S0006-3495(92)81572-2} {\bibfield  {journal}
  {\bibinfo  {journal} {Biophys. J.}\ }\textbf {\bibinfo {volume} {63}},\
  \bibinfo {pages} {98--110} (\bibinfo {year} {1992})}\BibitemShut {NoStop}%
\bibitem [{\citenamefont {Wofsy}\ and\ \citenamefont
  {Goldstein}(1992)}]{wofsy1992b}%
  \BibitemOpen
  \bibfield  {author} {\bibinfo {author} {\bibfnamefont {C.}~\bibnamefont
  {Wofsy}}\ and\ \bibinfo {author} {\bibfnamefont {B.}~\bibnamefont
  {Goldstein}},\ }\bibfield  {title} {\bibinfo {title} {{Interpretation of
  Scatchard plots for aggregating receptor systems.}},\ }\href
  {https://doi.org/10.1016/0025-5564(92)90090-J} {\bibfield  {journal}
  {\bibinfo  {journal} {Math. Biosci.}\ }\textbf {\bibinfo {volume} {112}},\
  \bibinfo {pages} {115--54} (\bibinfo {year} {1992})}\BibitemShut {NoStop}%
\bibitem [{\citenamefont {Klein}\ \emph {et~al.}(2004)\citenamefont {Klein},
  \citenamefont {Mattoon}, \citenamefont {Lemmon},\ and\ \citenamefont
  {Schlessinger}}]{klein2004}%
  \BibitemOpen
  \bibfield  {author} {\bibinfo {author} {\bibfnamefont {P.}~\bibnamefont
  {Klein}}, \bibinfo {author} {\bibfnamefont {D.}~\bibnamefont {Mattoon}},
  \bibinfo {author} {\bibfnamefont {M.~A.}\ \bibnamefont {Lemmon}},\ and\
  \bibinfo {author} {\bibfnamefont {J.}~\bibnamefont {Schlessinger}},\
  }\bibfield  {title} {\bibinfo {title} {{A structure-based model for ligand
  binding and dimerization of EGF receptors}},\ }\href@noop {} {\bibfield
  {journal} {\bibinfo  {journal} {Proc. Natl. Acad. Sci. U.S.A.}\ }\textbf
  {\bibinfo {volume} {101}} (\bibinfo {year} {2004})}\BibitemShut {NoStop}%
\bibitem [{\citenamefont {Mayawala}\ \emph
  {et~al.}(2005{\natexlab{a}})\citenamefont {Mayawala}, \citenamefont
  {Vlachos},\ and\ \citenamefont {Edwards}}]{mayawala2005a}%
  \BibitemOpen
  \bibfield  {author} {\bibinfo {author} {\bibfnamefont {K.}~\bibnamefont
  {Mayawala}}, \bibinfo {author} {\bibfnamefont {D.~G.}\ \bibnamefont
  {Vlachos}},\ and\ \bibinfo {author} {\bibfnamefont {J.~S.}\ \bibnamefont
  {Edwards}},\ }\bibfield  {title} {\bibinfo {title} {{Computational modeling
  reveals molecular details of epidermal growth factor binding}},\ }\href
  {https://doi.org/10.1186/1471-2121-6-41} {\bibfield  {journal} {\bibinfo
  {journal} {BMC Cell Biol.}\ }\textbf {\bibinfo {volume} {6}},\ \bibinfo
  {pages} {1--11} (\bibinfo {year} {2005}{\natexlab{a}})}\BibitemShut {NoStop}%
\bibitem [{\citenamefont {Mayawala}\ \emph
  {et~al.}(2005{\natexlab{b}})\citenamefont {Mayawala}, \citenamefont
  {Vlachos},\ and\ \citenamefont {Edwards}}]{mayawala2005b}%
  \BibitemOpen
  \bibfield  {author} {\bibinfo {author} {\bibfnamefont {K.}~\bibnamefont
  {Mayawala}}, \bibinfo {author} {\bibfnamefont {D.~G.}\ \bibnamefont
  {Vlachos}},\ and\ \bibinfo {author} {\bibfnamefont {J.~S.}\ \bibnamefont
  {Edwards}},\ }\bibfield  {title} {\bibinfo {title} {{Heterogeneities in EGF
  receptor density at the cell surface can lead to concave up scatchard plot of
  EGF binding}},\ }\href {https://doi.org/10.1016/j.febslet.2005.04.059}
  {\bibfield  {journal} {\bibinfo  {journal} {FEBS Lett.}\ }\textbf {\bibinfo
  {volume} {579}},\ \bibinfo {pages} {3043--3047} (\bibinfo {year}
  {2005}{\natexlab{b}})}\BibitemShut {NoStop}%
\bibitem [{\citenamefont {Ozcan}\ \emph {et~al.}(2006)\citenamefont {Ozcan},
  \citenamefont {Klein}, \citenamefont {Lemmon}, \citenamefont {Lax},\ and\
  \citenamefont {Schlessinger}}]{ozcan2006}%
  \BibitemOpen
  \bibfield  {author} {\bibinfo {author} {\bibfnamefont {F.}~\bibnamefont
  {Ozcan}}, \bibinfo {author} {\bibfnamefont {P.}~\bibnamefont {Klein}},
  \bibinfo {author} {\bibfnamefont {M.~a.}\ \bibnamefont {Lemmon}}, \bibinfo
  {author} {\bibfnamefont {I.}~\bibnamefont {Lax}},\ and\ \bibinfo {author}
  {\bibfnamefont {J.}~\bibnamefont {Schlessinger}},\ }\bibfield  {title}
  {\bibinfo {title} {{On the nature of low- and high-affinity EGF receptors on
  living cells}},\ }\href {https://doi.org/10.1073/pnas.0601469103} {\bibfield
  {journal} {\bibinfo  {journal} {Proc. Natl. Acad. Sci. U.S.A.}\ }\textbf
  {\bibinfo {volume} {103}},\ \bibinfo {pages} {5735--5740} (\bibinfo {year}
  {2006})}\BibitemShut {NoStop}%
\bibitem [{\citenamefont {Macdonald}\ and\ \citenamefont
  {Pike}(2008)}]{macdonald2008}%
  \BibitemOpen
  \bibfield  {author} {\bibinfo {author} {\bibfnamefont {J.~L.}\ \bibnamefont
  {Macdonald}}\ and\ \bibinfo {author} {\bibfnamefont {L.~J.}\ \bibnamefont
  {Pike}},\ }\bibfield  {title} {\bibinfo {title} {{Heterogeneity in
  EGF-binding affinities arises from negative cooperativity in an aggregating
  system}},\ }\href {https://doi.org/10.1073/pnas.0710524} {\bibfield
  {journal} {\bibinfo  {journal} {Proc. Natl. Acad. Sci. U.S.A.}\ }\textbf
  {\bibinfo {volume} {105}},\ \bibinfo {pages} {20147--20148} (\bibinfo {year}
  {2008})}\BibitemShut {NoStop}%
\bibitem [{\citenamefont {Lemmon}(2008)}]{lemmon2008}%
  \BibitemOpen
  \bibfield  {author} {\bibinfo {author} {\bibfnamefont {M.~A.}\ \bibnamefont
  {Lemmon}},\ }\bibfield  {title} {\bibinfo {title} {{Ligand-induced ErbB
  receptor dimerization}},\ }\href
  {https://doi.org/10.1016/j.yexcr.2008.10.024} {\bibfield  {journal} {\bibinfo
   {journal} {Experimental Cell Research}\ }\textbf {\bibinfo {volume} {315}},\
  \bibinfo {pages} {638--648} (\bibinfo {year} {2008})}\BibitemShut {NoStop}%
\bibitem [{\citenamefont {Adak}\ \emph {et~al.}(2011)\citenamefont {Adak},
  \citenamefont {DeAndrade},\ and\ \citenamefont {Pike}}]{adak2011}%
  \BibitemOpen
  \bibfield  {author} {\bibinfo {author} {\bibfnamefont {S.}~\bibnamefont
  {Adak}}, \bibinfo {author} {\bibfnamefont {D.}~\bibnamefont {DeAndrade}},\
  and\ \bibinfo {author} {\bibfnamefont {L.~J.}\ \bibnamefont {Pike}},\
  }\bibfield  {title} {\bibinfo {title} {{The tethering arm of the EGF receptor
  is required for negative cooperativity and signal transduction}},\ }\href
  {https://doi.org/10.1074/jbc.M110.182899} {\bibfield  {journal} {\bibinfo
  {journal} {J. Biol. Chem.}\ }\textbf {\bibinfo {volume} {286}},\ \bibinfo
  {pages} {1545--1555} (\bibinfo {year} {2011})}\BibitemShut {NoStop}%
\bibitem [{\citenamefont {Pike}(2012)}]{pike2012}%
  \BibitemOpen
  \bibfield  {author} {\bibinfo {author} {\bibfnamefont {L.}~\bibnamefont
  {Pike}},\ }\bibfield  {title} {\bibinfo {title} {{Negative co-operativity in
  the EGF receptor}},\ }\href {https://doi.org/10.1042/BST20110610} {\bibfield
  {journal} {\bibinfo  {journal} {Biochem. Soc. Trans.}\ }\textbf {\bibinfo
  {volume} {40}},\ \bibinfo {pages} {15--19} (\bibinfo {year}
  {2012})}\BibitemShut {NoStop}%
\bibitem [{\citenamefont {Hiroshima}\ \emph {et~al.}(2012)\citenamefont
  {Hiroshima}, \citenamefont {Saeki}, \citenamefont {Okada-Hatakeyama},\ and\
  \citenamefont {Sako}}]{hiroshima2012}%
  \BibitemOpen
  \bibfield  {author} {\bibinfo {author} {\bibfnamefont {M.}~\bibnamefont
  {Hiroshima}}, \bibinfo {author} {\bibfnamefont {Y.}~\bibnamefont {Saeki}},
  \bibinfo {author} {\bibfnamefont {M.}~\bibnamefont {Okada-Hatakeyama}},\ and\
  \bibinfo {author} {\bibfnamefont {Y.}~\bibnamefont {Sako}},\ }\bibfield
  {title} {\bibinfo {title} {{Dynamically varying interactions between
  heregulin and ErbB proteins detected by single-molecule analysis in living
  cells.}},\ }\href {https://doi.org/10.1073/pnas.1200464109} {\bibfield
  {journal} {\bibinfo  {journal} {Proc. Natl. Acad. Sci. U.S.A.}\ }\textbf
  {\bibinfo {volume} {109}},\ \bibinfo {pages} {13984--9} (\bibinfo {year}
  {2012})}\BibitemShut {NoStop}%
\bibitem [{\citenamefont {Hiroshima}\ and\ \citenamefont
  {Sako}(2013)}]{hiroshima2013}%
  \BibitemOpen
  \bibfield  {author} {\bibinfo {author} {\bibfnamefont {M.}~\bibnamefont
  {Hiroshima}}\ and\ \bibinfo {author} {\bibfnamefont {Y.}~\bibnamefont
  {Sako}},\ }\bibfield  {title} {\bibinfo {title} {{Regulation Mechanism of
  ErbB-Heregulin Interaction Shown by Single-molecule Kinetic Analysis in
  Living Cells.}},\ }\href@noop {} {\bibfield  {journal} {\bibinfo  {journal}
  {Biophys. Physicobiol.}\ }\textbf {\bibinfo {volume} {53}},\ \bibinfo {pages}
  {317--318} (\bibinfo {year} {2013})}\BibitemShut {NoStop}%
\bibitem [{\citenamefont {Tomita}\ \emph {et~al.}(1999)\citenamefont {Tomita},
  \citenamefont {Hashimoto}, \citenamefont {Takahashi}, \citenamefont
  {Shimizu}, \citenamefont {Matsuzaki}, \citenamefont {Miyoshi}, \citenamefont
  {Saito}, \citenamefont {Tanida}, \citenamefont {Yugi},\ and\ \citenamefont
  {Venter}}]{tomita1999}%
  \BibitemOpen
  \bibfield  {author} {\bibinfo {author} {\bibfnamefont {M.}~\bibnamefont
  {Tomita}}, \bibinfo {author} {\bibfnamefont {K.}~\bibnamefont {Hashimoto}},
  \bibinfo {author} {\bibfnamefont {K.}~\bibnamefont {Takahashi}}, \bibinfo
  {author} {\bibfnamefont {T.}~\bibnamefont {Shimizu}}, \bibinfo {author}
  {\bibfnamefont {Y.}~\bibnamefont {Matsuzaki}}, \bibinfo {author}
  {\bibfnamefont {F.}~\bibnamefont {Miyoshi}}, \bibinfo {author} {\bibfnamefont
  {K.}~\bibnamefont {Saito}}, \bibinfo {author} {\bibfnamefont
  {S.}~\bibnamefont {Tanida}}, \bibinfo {author} {\bibfnamefont
  {K.}~\bibnamefont {Yugi}},\ and\ \bibinfo {author} {\bibfnamefont
  {J.}~\bibnamefont {Venter}},\ }\bibfield  {title} {\bibinfo {title} {{E-CELL:
  Software environment for whole-cell simulation.}},\ }\href@noop {} {\bibfield
   {journal} {\bibinfo  {journal} {Bioinformatics}\ }\textbf {\bibinfo {volume}
  {15}},\ \bibinfo {pages} {72--84} (\bibinfo {year} {1999})}\BibitemShut
  {NoStop}%
\bibitem [{\citenamefont {Arjunan}\ and\ \citenamefont
  {Tomita}(2010)}]{arjunan2010}%
  \BibitemOpen
  \bibfield  {author} {\bibinfo {author} {\bibfnamefont {S.~N.~V.}\
  \bibnamefont {Arjunan}}\ and\ \bibinfo {author} {\bibfnamefont
  {M.}~\bibnamefont {Tomita}},\ }\bibfield  {title} {\bibinfo {title} {{A new
  multicompartmental reaction-diffusion modeling method links transient
  membrane attachment of E. coli MinE to E-ring formation}},\ }\href
  {https://doi.org/10.1007/s11693-009-9047-2} {\bibfield  {journal} {\bibinfo
  {journal} {Syst. Synth. Biol.}\ }\textbf {\bibinfo {volume} {4}},\ \bibinfo
  {pages} {35--53} (\bibinfo {year} {2010})}\BibitemShut {NoStop}%
\bibitem [{\citenamefont {Itano}\ \emph {et~al.}(2018)\citenamefont {Itano},
  \citenamefont {Ito}, \citenamefont {Kawasaki}, \citenamefont {Murakami},\
  and\ \citenamefont {Suzuki}}]{itano2018}%
  \BibitemOpen
  \bibfield  {author} {\bibinfo {author} {\bibfnamefont {K.}~\bibnamefont
  {Itano}}, \bibinfo {author} {\bibfnamefont {T.}~\bibnamefont {Ito}}, \bibinfo
  {author} {\bibfnamefont {S.}~\bibnamefont {Kawasaki}}, \bibinfo {author}
  {\bibfnamefont {Y.}~\bibnamefont {Murakami}},\ and\ \bibinfo {author}
  {\bibfnamefont {T.}~\bibnamefont {Suzuki}},\ }\bibfield  {title} {\bibinfo
  {title} {{Mathematical modeling and analysis of ErbB3 and EGFR dimerization
  process for the gefitinib resistance}},\ }\href
  {https://doi.org/10.14495/jsiaml.10.33} {\bibfield  {journal} {\bibinfo
  {journal} {JSIM Letters}\ }\textbf {\bibinfo {volume} {10}},\ \bibinfo
  {pages} {33--36} (\bibinfo {year} {2018})}\BibitemShut {NoStop}%
\bibitem [{\citenamefont {Ito}\ \emph {et~al.}(2019)\citenamefont {Ito},
  \citenamefont {Kumagai}, \citenamefont {Itano}, \citenamefont {Maruyama},
  \citenamefont {Tamura}, \citenamefont {Kawasaki}, \citenamefont {Suzuki},\
  and\ \citenamefont {Murakami}}]{ito2019}%
  \BibitemOpen
  \bibfield  {author} {\bibinfo {author} {\bibfnamefont {T.}~\bibnamefont
  {Ito}}, \bibinfo {author} {\bibfnamefont {Y.}~\bibnamefont {Kumagai}},
  \bibinfo {author} {\bibfnamefont {K.}~\bibnamefont {Itano}}, \bibinfo
  {author} {\bibfnamefont {T.}~\bibnamefont {Maruyama}}, \bibinfo {author}
  {\bibfnamefont {K.}~\bibnamefont {Tamura}}, \bibinfo {author} {\bibfnamefont
  {S.}~\bibnamefont {Kawasaki}}, \bibinfo {author} {\bibfnamefont
  {T.}~\bibnamefont {Suzuki}},\ and\ \bibinfo {author} {\bibfnamefont
  {Y.}~\bibnamefont {Murakami}},\ }\bibfield  {title} {\bibinfo {title}
  {{Mathematical analysis of ge fi tinib resistance of lung adenocarcinoma
  caused by MET ampli fi cation}},\ }\href
  {https://doi.org/10.1016/j.bbrc.2019.02.086} {\bibfield  {journal} {\bibinfo
  {journal} {Biochem. Biophys. Res. Commun.}\ }\textbf {\bibinfo {volume}
  {511}},\ \bibinfo {pages} {544--550} (\bibinfo {year} {2019})}\BibitemShut
  {NoStop}%
\end{thebibliography}


\end{document}